\documentclass[10pt]{article}
\usepackage{graphicx}
\usepackage{amsmath,amssymb}
\newcommand{\rmd}{\mathrm{d}}
\newcommand{\rmi}{\mathrm{i}}
\newcommand{\bi}[1]{\textbf{\textit{#1}}}

\begin{document}
\title{\textbf{Effects of environmental and exciton screening in single-walled carbon nanotubes}}
\author{Vadym M. Adamyan\footnote{E-mail: vadamyan@paco.net}, Oleksii A. Smyrnov\footnote{E-mail: smyrnov@onu.edu.ua}, Sergey V. Tishchenko\\
\emph{Department of Theoretical Physics, Odessa I. I. Mechnikov
National University,}\\ \emph{2 Dvoryanskaya St., Odessa 65026,
Ukraine}}
\date{July 19, 2008}
\maketitle
\begin{abstract}
The ground-state exciton binding energy for single-walled carbon
nanotubes (SWCNTs) in vacuum calculated ignoring the screening of
Coulomb interaction appears to be much greater than the
corresponding band gap. The most essential contributions to the
screening of electron-hole (e-h) interaction potential in
semiconducting SWCNTs, which return the ground-state exciton
binding energy into the energy gap, are considered. Our estimates
on the screening effects and exciton binding energies are in
satisfactory agreement with the corresponding experimental data
for concrete nanotubes.
\end{abstract}

PACS number(s): 78.67.Ch

\section{Introduction}
\setcounter{equation}{0}

As it was shown in the recent works on optical spectra of SWCNTs,
both theoretical~\cite{ando}-\cite{jdd} and
experimental~\cite{bachilo}-\cite{kiow}, the study of excitonic
contributions in those spectra is of prime importance. Evidently,
the quasione-dimensional large-radius exciton problem can be
reduced to the 1D system of two quasi-particles with the potential
having Coulomb attraction tail. Indeed, within the framework of
the so-called long-wave approximation the wave equation for the
envelope function Fourier transform $\phi$ of a large-radius rest
exciton in a semiconducting SWCNT with the longitudinal period $a$
is reduced to the following 1D Schr\"{o}dinger equation:
\begin{equation}\label{1.1}
-\frac{\hbar^2}{2\mu}\phi''(z)+V(z)\phi(z)=\mathcal{E}\phi(z),\
\mathcal{E}=E_\mathrm{exc}-E_\mathrm{g}, \  -\infty<z<\infty,
\end{equation}
with the exciton reduced effective mass $\mu$ and the e-h
interaction potential
\begin{equation}\label{1.2}
V(z)=-\int\limits_{\mathrm{E}_3^a}\int\limits_{\mathrm{E}_3^a}\frac{e^2|u_{\mathrm{c};0}(\bi{r}_1)|^2|u_{\mathrm{v};0}(\bi{r}_2)|^2\rmd\bi{r}_1\rmd\bi{r}_2}{\left((x_1-x_2)^2+(y_1-y_2)^2+(z+z_1-z_2)^2\right)^{1/2}},\
\ \mathrm{E}_3^a=\mathrm{E}_2\times(0<z<a).
\end{equation}
Here $u_{\mathrm{c,v};q}(\bi{r})$ are the Bloch amplitudes of the
Bloch wave functions $\psi_{\mathrm{c,v};q}(\bi{r})=\exp(\rmi
qz)u_{\mathrm{c,v};q}(\bi{r})$ of the conduction and valence band
electrons of a SWCNT, respectively, $q$ is the electron
quasi-momentum. Under the assumption that the charges of electron
and hole participating in the formation of exciton are smeared
uniformly along infinitesimal narrow bands at the nanotube wall we
obtain from~(\ref{1.2}):
\begin{equation}\label{1.3}
V_{R_0}(z)=-\frac{e^2}{4\pi^2|z|}\int\limits_0^{2\pi}\int\limits_0^{2\pi}
\frac{\rmd\alpha_1\rmd\alpha_2}{\left(1+(4R_0^2/z^2)\sin^2{\frac{\alpha_1-\alpha_2}{2}}\right)^{1/2}}=-\frac{2e^2}{\pi
|z|}\mathrm{K}\left[-\frac{4R^2_0}{z^2}\right],
\end{equation}
where $\mathrm{K}$ is the complete elliptic integral of the first
kind and $R_0$ is the tube radius. This potential is the simplest
approximation to the bare Coulomb potential, which accounts the
finiteness of the nanotube diameter.

Due to the parity of the interaction potential the exciton states
should split into the odd and even series. In~\cite{adsmyr} we
show that for the bare e-h interaction potentials $V(z)=-e^2/|z|$
and~(\ref{1.3}), and for the e-h interaction potentials screened
by the nanotube bound electrons or by free charges, which may
appear in a SWCNT at rather high temperature, the binding energy
of even excitons for SWCNTs in vacuum in the ground state well
exceeds the energy gap. This may lead to instability of the
single-electron states at least in the vicinity of the energy gap
with respect to formation of excitons. In~\cite{adsmyr} we also
briefly discussed the factors, which could prevent the collapse of
single-electron states in isolated semiconducting SWCNTs. Here we
present more detailed analysis of those factors (with the
corresponding calculations), namely we consider: the environmental
screening of e-h interaction for nanotubes in a medium (section~2)
and the screening by excitons, which appear in a nanotube in
vacuum due to the mentioned instability of single-electron states
(section~3). These screening effects substantially weaken the e-h
interaction and return the exciton ground-state binding energy
into the corresponding band gap. Particularly, the appearance of a
rather small concentration of excitons  stabilizes the
single-electron states in SWCNT preventing their further
conversion into excitons. The obtained data and estimates were
compared with the corresponding experimental results (section~4).

\section{Environmental screening}
\setcounter{equation}{0}

Evidently, a dielectric medium, surrounding a nanotube, should
substantially change the e-h interaction potential. In
experimental works~\cite{bachilo}-\cite{wang} (which used the
methods described in~\cite{fluor}) investigated individual
nanotubes were not in vacuum but encased in sodium dodecyl sulfate
(SDS) cylindrical micelles disposed in $\mathrm{D_2O}$. Because of
these SDS micelles, which provided a pure hydrocarbon environment
around individual nanotubes, the high permittivity solvent
$\mathrm{D_2O}$ did not reach nanotubes. However, the environment
of hydrophobic hydrocarbon "tails" $(-\mathrm{C_{12}H_{25}})$ of
the SDS molecules has the permittivity greater than unity
(according to~\cite{kiow} it's about 2-2.5). Following the
figure~1A from~\cite{fluor} we considered a simple model of a
SWCNT in a dielectric environment: a narrow, infinite cylinder
with radius $R_0$ in a medium with the dielectric constant
$\varepsilon_\mathrm{env}$ and the internal dielectric constant
$\varepsilon_\mathrm{int}$.

Let us find a screened analogue of potential~(\ref{1.3}) for this
model under the assumption about axially symmetrical charge
localization at nanotube's (cylinder's) wall. To obtain the sought
screened potential $\varphi$ we consider as in~\cite{banyai} the
following boundary problem for the one-dimensional Fourier
transform of the Laplace equation:
\begin{equation}\label{2.1}
-\Delta_\mathrm{2D}\varphi(k,\bi{r}_\mathrm{2D})+k^2\varphi(k,\bi{r}_\mathrm{2D})=0,
\  r_\mathrm{2D}\lessgtr
R_0,~\varphi(k,0)<\infty,~\varphi(k,\infty)=0
\end{equation}
and two standard boundary conditions for the potential at the tube
surface:
\begin{equation}\label{2.2}
\begin{split}
&\varphi(z,R_0-0)=\varphi(z,R_0+0),\\
&\varepsilon_\mathrm{env}\left.\frac{\partial\varphi(z,\bi{r}_\mathrm{2D})}{\partial
r_\mathrm{2D}}\right|_{r_\mathrm{2D}=R_0+0}-\varepsilon_\mathrm{int}\left.\frac{\partial\varphi(z,\bi{r}_\mathrm{2D})}{\partial
r_\mathrm{2D}}\right|_{r_\mathrm{2D}=R_0-0}=4\pi\sigma(z),
\end{split}
\end{equation}
where $\bi{r}_\mathrm{2D}$ is the transverse component of the
radius-vector and $\sigma=(e/2\pi R_0)\delta(z)$ is the surface
density of the screened charge distribution. Due to the axial
symmetry of charges distribution the differential equation
in~(\ref{2.1}) can be reduced to the modified Bessel equation
(with different solutions for $r_\mathrm{2D}<R_0$ and
$r_\mathrm{2D}>R_0$), from which one can simply obtain the 1D
screened potential:
\begin{equation}\label{2.4}
\varphi(z,R_0)=-\frac{2e}{\pi
R_0}\int\limits^\infty_0\frac{I_0(k)K_0(k)\cos(kz/R_0)}{[\varepsilon_\mathrm{env}
I_0(k)K_1(k)+\varepsilon_\mathrm{int}I_1(k)K_0(k)]k}\rmd k,
\end{equation}
where $I_i$ and $K_i$ are the modified Bessel functions of the
order $i$ of the first and second kind, respectively. For a
nanotube in medium internal screening is mainly induced by the
nanotube $\pi$-electrons and in this case we take
$\varepsilon_\mathrm{int}(k)$ obtained in~\cite{adsmyr} in
section~3.

\section{Screening induced by excitons}
\setcounter{equation}{0}

As it was mentioned above, since the large exciton ground-state
binding energy exceeds the corresponding energy gap, then the
single-electron states in any nanotube in vacuum
($\varepsilon_\mathrm{env}=1$) should be unstable with respect to
the formation of excitons. But with the advent of some number of
excitons in the tube the additional screening effect, stipulated
by a rather great polarizability of excitons in the longitudinal
electric field, appears. Under certain critical concentration of
excitons the ground-state exciton binding energy becomes smaller
than the energy gap and the conversion of single-electron states
into excitons ends. Hence, here we'll obtain the upper and lower
limits for the critical concentration of excitons.

As is well known, the permittivity of any dielectric, and so the
permittivity of the exciton gas, can be given as follows:
\begin{equation}\label{3.1}
\varepsilon_\mathrm{exc}=1+4\pi\alpha,\ \ \
\alpha=2e^2n\sum\limits_k\frac{|\langle\Psi_0|\bi{r}|\Psi_k\rangle|^2}{\mathcal{E}_0-\mathcal{E}_k},
\end{equation}
where $\alpha$ is the polarizability of the exciton gas in the
static electric field, $n$ is the bulk concentration of excitons,
$\Psi_0$ and $\mathcal{E}_0$ are the exciton eigenfunction and
binding energy, which correspond to the ground state, and $\Psi_k$
and $\mathcal{E}_k$ are those, which correspond to the all excited
states of exciton. According to~(\ref{3.1}) the upper and lower
limits for $\alpha$ are:
\begin{equation}\label{3.2}
\frac{2e^2n}{\mathcal{E}_0-\mathcal{E}_1}|\langle\Psi_0|\bi{r}|\Psi_1\rangle|^2\leq\alpha\leq\frac{2e^2n}{\mathcal{E}_0-\mathcal{E}_1}\sum\limits_k|\langle\Psi_0|\bi{r}|\Psi_k\rangle|^2=\frac{2e^2n}{\mathcal{E}_0-\mathcal{E}_1}|\langle\Psi_0|\bi{r}^2|\Psi_0\rangle|,
\end{equation}
where $\Psi_1$ and $\mathcal{E}_1$ correspond to the lowest
excited exciton state. Further, assuming that the external
electric field is longitudinal (directed along the $z$-axis,
nanotube axis) and simplifying matrix elements in~(\ref{3.2}), in
the long-wave limit we can write the upper and lower limits for
the concentration of excitons in the following form:
\begin{equation}\label{3.3}
\frac{\varepsilon_\mathrm{exc}-1}{4\pi}\frac{\mathcal{E}_0-\mathcal{E}_1}{2e^2}\left|
\int\limits^\infty_{-\infty}z^2|\phi_0(z)|^2\rmd z\right|^{-1}\leq
n\leq\frac{\varepsilon_\mathrm{exc}-1}{4\pi}\frac{\mathcal{E}_0-\mathcal{E}_1}{2e^2}\left|
\int\limits^{\infty}_{-\infty}z\phi_0(z)\phi_1(z)\rmd
z\right|^{-2},
\end{equation}
where each $\phi$ is the component of Fourier transform of the
corresponding exciton envelope function, it depends only on the
distance $z$ between the electron and hole. Each $\phi$ is the
solution of wave equation~(\ref{1.1}) with potential~(\ref{2.4}),
where $\varepsilon_\mathrm{env}=1$ and
$\varepsilon_\mathrm{int}=\varepsilon_\mathrm{exc}=\mathrm{const}$
as the screening induced by the nanotube bound electrons is
negligible in comparison with $\varepsilon_\mathrm{exc}$. $\phi_0$
is the even function, which corresponds to the exciton ground
state and satisfies the 1D Schr\"{o}dinger equation~(\ref{1.1})
and the boundary condition at the origin $\phi'(0)=0$, and
$\phi_1$ is the odd function, which corresponds to the lowest
excited exciton state and satisfies the same equation, but the
boundary condition $\phi(0)=0$ at the origin.

Varying $\varepsilon_\mathrm{exc}$ in~(\ref{2.4}) substituted into
wave equation~(\ref{1.1}) one can match $\mathcal{E}_0$ to the
energy gap. Further, $\mathcal{E}_1$ can be obtained from the same
equation with the fixed $\varepsilon_\mathrm{exc}$ and with the
corresponding boundary condition. These magnitudes allow to
calculate from~(\ref{3.3}) the rough upper and lower limits for
the critical concentration of excitons $n_\mathrm{c}$. Further,
knowing $n_\mathrm{c}$ we can calculate the shift of the forbidden
band edges, which move apart due to the transformation of some
single-electron states into excitons. This results in the
enhancement of energy gap and hence the lowest optical transition
energy $E_{11}$ should be blueshifted by
\begin{equation}\label{3.4}
\delta
E_{11}=\left(\frac{1}{m^*_\mathrm{e}}+\frac{1}{m^*_\mathrm{h}}\right)\frac{(\pi\hbar\widetilde{n}_\mathrm{c})^2}{2}
\end{equation}
as in~\cite{ohno} and~\cite{kiow}. Here
$\widetilde{n}_\mathrm{c}=n_\mathrm{c}\pi R_0^2$ is the linear
critical concentration of excitons and $m^*_\mathrm{e,h}$ are the
electrons and holes effective masses ($m^*_\mathrm{e}\simeq
m^*_\mathrm{h}$ for all SWCNTs).

\section{Calculation results. Discussion}
\setcounter{equation}{0}

Parameters of electronic structure of nanotubes used in this work,
were calculated as in~\cite{tish},~\cite{adtish} within the
framework of the zero-range potentials
method~\cite{aghh},~\cite{demk}.

Using potential~(\ref{2.4}) we've calculated the exciton binding
energy for nanotube (7,~5) ($R_0=0.4087~\mathrm{nm}$, band gap
$E_\mathrm{g}=1.01~\mathrm{eV}$) in the SDS hydrocarbon medium
(by~\cite{kiow} its dielectric constant
$\varepsilon_\mathrm{env}=2\div2.5$). The experimental value of
the exciton binding energy for (7,~5) tube encased in the SDS
micelle is about $0.62\pm0.05~\mathrm{eV}$ ~\cite{wang}. Our
result in this case is $0.90\div0.68~\mathrm{eV}$. Recall, that
in~\cite{cap} this binding energy was obtained using
$\varepsilon=2.559$.

The experimental value of blueshift for the (7,~5) SWCNT is
$\delta E_{11}=40-55~\mathrm{meV}$~\cite{kiow}. By~(\ref{3.4}) for
the (7,~5) tube this gives
$\widetilde{n}_\mathrm{c}\sim100~\mu\mathrm{m}^{-1}$, while from
estimates~(\ref{3.3}) it follows that
$\widetilde{n}_\mathrm{c}\in[150,160]~\mu\mathrm{m}^{-1}$. The
discrepancy may be stipulated by ignoring of the collective
effects in exciton gas and by using of the Schr\"{o}dinger
equation near the band edge instead of the Bethe-Salpeter equation
with the energy-dependent potential~\cite{spb}.

The exciton binding energy was also calculated for the (8,~0)
nanotube ($R_0=0.315~\mathrm{nm}$, band gap
$E_\mathrm{g}=1.415~\mathrm{eV}$). According to~(\ref{1.1})
with~(\ref{2.4}) we have $1.26\div0.97~\mathrm{eV}$, which is
close to that in~\cite{spataru} (about $0.86-1~\mathrm{eV}$).
Recall that the results on the (8,~0) nanotube in~\cite{spataru}
are in good agreement with those obtained in~\cite{bachilo2} by
interpolation of experimental data for another nanotubes in SDS
micelles.

\end{document}